\documentclass[pre,amsmath,amssymb,twocolumn,superscriptaddress,showpacs]{revtex4-1}

\pdfoutput=1 

\usepackage{graphicx}
\usepackage{dcolumn}
\usepackage{bm}

\begin{document}

\newcommand{\bea}{\begin{eqnarray}}
\newcommand{\eea}{  \end{eqnarray}}
\newcommand{\bit}{\begin{itemize}}
\newcommand{\eit}{  \end{itemize}}

\newcommand{\be}{\begin{equation}}
\newcommand{\ee}{\end{equation}}
\newcommand{\ra}{\rangle}
\newcommand{\la}{\langle}
\newcommand{\U}{\widetilde{U}} 
\newcommand{\dpc}[1]{\textcolor{red}{(DP: #1)}}
\newcommand{\ntxt}[1]{\textcolor{blue}{#1}} 
\newcommand{\rhop}{\hat{\rho}} 
\newcommand{\he}{\hbar_{\rm eff}}   


\def\bra#1{{\langle#1|}}
\def\ket#1{{|#1\rangle}}
\def\bracket#1#2{{\langle#1|#2\rangle}}
\def\inner#1#2{{\langle#1|#2\rangle}}
\def\expect#1{{\langle#1\rangle}}
\def\e{{\rm e}}
\def\proj{{\hat{\cal P}}}
\def\tr{{\rm Tr}}
\def\H{{\hat H}}
\def\Hdag{{\hat H}^\dagger}
\def\Lop{{\cal L}}
\def\Ehat{{\hat E}}
\def\Edag{{\hat E}^\dagger}
\def\Shat{\hat{S}}
\def\Sdag{{\hat S}^\dagger}
\def\Ahat{{\hat A}}
\def\Adag{{\hat A}^\dagger}
\def\U{{\hat U}}
\def\Udag{{\hat U}^\dagger}
\def\Zhat{{\hat Z}}
\def\Phat{{\hat P}}
\def\Op{{\hat O}}
\def\id{{\hat I}}
\def\x{{\hat x}}
\def\P{{\hat P}}
\def\Px{\proj_x}
\def\Pr{\proj_{R}}
\def\Pl{\proj_{L}}


\title{Classical counterparts of quantum attractors in generic dissipative systems}

\author{Gabriel G. Carlo}
\email{carlo@tandar.cnea.gov.ar}
\author{Leonardo Ermann}
\author{Alejandro M. F. Rivas}
\author{Mar\'\i a E. Spina}
\affiliation{Departamento de F\'\i sica, CNEA, Libertador 8250, (C1429BNP) Buenos Aires, Argentina}
\author{Dario Poletti}
\affiliation{Singapore University of Technology and Design, 8 Somapah Road, 
487372 Singapore}

\date{\today}

\pacs{05.45.Mt, 03.65.Yz, 05.45.−a}

\begin{abstract}

In the context of dissipative systems, we show that for any quantum chaotic attractor a corresponding 
classical chaotic attractor can always be found. We provide with a general way to locate them, rooted in 
the structure of the parameter space (which is typically bidimensional, accounting for the forcing strength and dissipation parameters). 
In the cases where an approximate point like quantum distribution is found, it can be associated to exceptionally 
large regular structures. Moreover, supposedly anomalous quantum chaotic behaviour can be very well reproduced by the classical dynamics plus 
Gaussian noise of the size of an effective Planck constant $\hbar_{\rm eff}$. We give support to our conjectures by means of two paradigmatic 
examples of quantum chaos and transport theory. In particular, a dissipative driven system becomes fundamental in order to extend 
their validity to generic cases.

\end{abstract}

\maketitle

\section{Introduction}
\label{sec1}

The study of the quantum to classical correspondence in dissipative systems is attracting a lot of attention 
nowadays. This is related to its relevance for many different fields that range from the theoretical aspects of 
quantum information \cite{Nielsen,Preskill} to applications such as in cold atoms \cite{CAexp1,CAexp2,AOKR1,AOKR2}. 
There is a new body of work that asks for a better understanding of the interplay between the classical and 
quantum properties of dissipation. We can mention the developments in reservoir engineering, which 
has been applied to generate robust quantum states in the presence of decoherence \cite{Kienzler}, for example. 
Optomechanics \cite{Bakemeier} is also revealing as a promising field where our knowledge of the many intricate features of the 
route to chaos, so deeply investigated in classical systems, needs to be extended to the quantum arena.  
Very recently, interesting properties of many body systems have been elucidated \cite{Hartmann,Ivanchenko}, and the corresponding 
classical equations could bring a whole new perspective when analyzed in terms of the signatures they 
imprint on the original quantum systems. A rocked open Bose-Hubbard dimer has shown 
a non trivial connection between the interactions and bifurcations in the mean field dynamics. Then, it is of 
the utmost importance to clarify any details that could be controversial. 

In this line, attention has been directed towards the effects of the monitoring (coupling) details on the emergence or 
suppression of chaos \cite{Eastman}. Moreover, an apparent paradox regarding regular quantum behaviour corresponding 
to a classical chaotic one in an optomechanical 
system has been nicely explained thanks to studies undertaken from the correspondence perspective \cite{Grebogi}. 
Finally, it was recently claimed that quantum chaotic attractors (i.e. complex quantum equilibrium states typically 
associated to classical chaotic attractors)
with no classical counterpart exist in the open dissipative quantum Duffing system \cite{Pokharel}.

In order to throw light over some of these features we study two paradigmatic 
systems: a dissipative modified kicked rotator map (DMKRM) which has been very fruitful in directed transport 
theory \cite{qdisratchets}, and a dissipative periodically driven dynamical system (DPDDS) that has applications 
in isomerization reactions \cite{isomerization} and which is fundamental to support the generic nature of our ideas. 
We concentrate on the case of a small but finite value of the effective Planck constant $\he$.
We have found that by suitably exploring the parameter space of these systems we are always able to find a 
classical chaotic attractor corresponding to any quantum chaotic one, even when the classical dynamics is 
regular (we propose a general way to do it). For the exceptional cases where no chaotic region is near 
the regular one in this space, the 
quantum limiting distributions become also regular (within quantum uncertainty). The addition of Gaussian noise of 
size $\hbar_{\rm eff}$ to the classical equations provides with the main features of the quantum evolution. The 
study of a generic system, beyond kicked ones, has been fundamental to extend the validity of these conjectures \cite{Carlo} 

We have organized our paper in the following way: In Sec. \ref{sec2} we describe the models including the 
way the extra Gaussian noise is added to the classical versions in order to find the quantum behaviour. The 
methods to integrate the equations are also explained. In Sec. \ref{sec3} we explore several key values of the 
parameters in order to show our main point, i.e. that a classical analog can always be found for 
quantum chaotic attractors. Also, we explain the exceptional cases where point like quantum distributions exist. 
In Sec. \ref{sec4} we conclude.

\section{Models and calculation methods}
\label{sec2}

\subsection{DMKRM}

The first model we consider is a particle moving in one dimension
[$x\in(-\infty,+\infty)$] periodically kicked by the potential:
\begin{equation}
V(x,t)=k\left[\cos(x)+\frac{a}{2}\cos(2x+\phi)\right]
\sum_{m=-\infty}^{+\infty}\delta(t-m \tau),
\end{equation}
where $k$ is the strength of each kick and $\tau$ is the kicking period. 
When adding dissipation we obtain the following map \cite{qdisratchets}
\begin{equation}
\left\{
\begin{array}{l}
\overline{n}=\gamma n +
k[\sin(x)+a\sin(2x+\phi)]
\\
\overline{x}=x+ \tau \overline{n}.
\end{array}
\right.
\label{dissmap}
\end{equation}
Here $n$ is the momentum variable conjugated to $x$ 
and $\gamma$ ($0\le \gamma \le 1$) is the dissipation parameter.
The conservative limit is reached at $\gamma=1$, whereas the 
value $\gamma=0$ gives the maximum damping. In order to simplify 
the parametric dependence it is usual to introduce a rescaled 
momentum variable $p=\tau n$ and the quantity $K=k \tau$. 
This is a paradigmatic model in directed transport. As such it shows a current that 
emerges as a consequence of breaking the spatial and temporal symmetries (when $a \neq 0$ with $\phi \neq m
\pi$, and  $\gamma \neq 1$). It is worth mentioning that we take $a=0.5$ 
and $\phi=\pi/2$ for this work.

Some of us have conjectured \cite{Carlo} that the main effects of the quantum 
fluctuations are similar to those of Gaussian fluctuations of the order of $\hbar_{\rm eff}$ 
induce in the classical map (we define the effective Planck constant $\hbar_{\rm eff}$ in the next paragraph). 
In order to introduce them we replace the first line of Eq. \ref{dissmap} with $\overline{n}=\gamma n +
k[\sin(x)+a\sin(2x+\phi)] + \xi$.  We have chosen to leave no free parameters 
in order to test the behaviour of our conjecture in this situation, so we fix $\langle \xi^2 \rangle = \hbar_{\rm eff}$, 
having zero mean. However, the exact coincidence of the size of fluctuations with $\hbar_{\rm eff}$ is not essential for it to be valid.

The quantum model (without noise) is obtained via:
$x\to \hat{x}$, $n\to \hat{n}=-i (d/dx)$ ($\hbar=1$).
Since $[\hat{x},\hat{p}]=i \tau$ (where $\hat{p}=\tau \hat{n}$), the effective Planck constant
is $\hbar_{\rm eff}=\tau$. The classical limit corresponds to
$\hbar_{\rm eff}\to 0$, while $K=\hbar_{\rm eff} k$ remains constant. 
Dissipation at the quantum level is introduced by means of the
master equation \cite{Lindblad} for the density operator $\hat{\rho}$ of the
system
\begin{equation}
\dot{\hat{\rho}} = -i
[\hat{H}_s,\hat{\rho}] - \frac{1}{2} \sum_{\mu=1}^2
\{\hat{L}_{\mu}^{\dag} \hat{L}_{\mu},\hat{\rho}\}+
\sum_{\mu=1}^2 \hat{L}_{\mu} \hat{\rho} \hat{L}_{\mu}^{\dag} \equiv \Lambda \rho.
\label{lindblad}
\end{equation}
Here $\hat{H}_s=\hat{n}^2/2+V(\hat{x},t)$ is the system
Hamiltonian, \{\,,\,\} is the anticommutator, and $\hat{L}_{\mu}$ are the Lindblad operators
given by \cite{Dittrich, Graham}
\begin{equation}
\begin{array}{l}
\hat{L}_1 = g \sum_n \sqrt{n+1} \; |n \rangle \, \langle n+1|,\\
\hat{L}_2 = g \sum_n \sqrt{n+1} \; |-n \rangle \, \langle -n-1|,
\end{array}
\end{equation}
with $n=0,1,...$ and $g=\sqrt{-\ln \gamma}$ to comply with the Ehrenfest theorem.

\subsection{DPDDS}

In order to give a more general support to our claims we have chosen to study a 
full dynamical system that can be thought as a particle moving in the continuously driven 
time periodic potential 

\begin{eqnarray}
  V(x,t) & = & 1- \cos(x)- A \cos(2x+\phi_a) +           \\ \nonumber
         &   & k \sin(x) \: [\cos(t)],
\end{eqnarray}

where $k$ is the strength of the time periodic forcing. 
Throughout this paper we take $A=0.5$ and $\phi_a=\pi/2$.
The picture is completed by means of a velocity dependent damping and Gaussian 
fluctuations that are usually taken as thermal ones, but that in our present study 
will play the same role as in the previous model, i.e. reproducing the main effects of 
quantum fluctuations.
Thus, we are led to numerically solve the equation

\begin{equation}
  m \ddot{x}=-\Gamma \dot{x} - V'(x,t) + \xi.
\end{equation}

As usual, $x$ stands for the spatial coordinate
of the particle, $m$ for its mass (we take $m=1$), and $\Gamma$ is the amount of dissipation.
Again, in a way similar to the DMKRM case, the Gaussian white noise having zero mean
$\xi$ is simply asked to satisfy $ \langle \xi(t) \xi(t') \rangle =\hbar_{\rm eff} \delta(t-t')$.

This system is interesting to simulate a molecule with 
two stable isomers that is under the influence of a monochromatic laser field pulse, 
for which the term $\sin(x)$ represents the dipole coupling \cite{isomerization}. 
But it is also of general nature, and the results obtained through its study can be 
directly applied to many different situations, including for example many body 
systems \cite{Hartmann}.

The quantum mechanical evolution is performed by means of a modified split operator method 
\cite{isomerization,SplitOp}.
In fact, we compose unitary steps given by the kinetic and potential
terms of the Hamiltonian, and other purely dissipative ones.
in order to treat these latter we use the same model as in the previous subsection. 
For the sake of completeness we explicitly write down the dissipative part of the Lindblad equation for the 
density matrix of this system, that can be written as a completely positive map 
${\bf D}_\alpha(dt)$ in the operator-sum or Kraus representation

\begin{eqnarray}
  \rho(t+dt) & = & {\bf D}_{(\varepsilon,T)}(dt)\left(\rho(t)\right)= \nonumber \\
             &   &  C_0 \rho(t) C^{\dagger}_0 + C^\pm_1 \rho(t) C^{\pm \dagger}_1,
  \label{krausrep}
\end{eqnarray}

where

\begin{eqnarray}
  C_0     & = & \openone-\frac{1}{2} \;
                {C^{\pm \dagger}_1 C^\pm_1} \nonumber \\
  C_1^\pm & = & \sum_{j} \sqrt{\varepsilon \;dt \; j}
                \; \ket{p_{\pm j \mp 1}}\bra{p_{\pm j}}  \nonumber \\
  \label{krauslindblad}
\end{eqnarray}

can be interpreted as infinitesimal Kraus operators obeying the rule 
$\sum_\mu{C^{\pm \dagger}_\mu C^\pm_\mu}=\openone$ to first order in $dt$ \cite{Carlo1}. 
We take $p$ as the momentum variable, conjugated to the $x$ coordinate.
In fact, the two different operators denoted by the superscript $\pm$ are associated to the positive
and negative values of the $p$ spectrum.
It is important to notice that in this case $\varepsilon$ is the system-bath coupling parameter and 
can be directly associated to the classical friction parameter $\Gamma$. 
Also, $\Gamma$ is taken differently from $\gamma$ in the DMKRM since in generic dynamical systems 
$\Gamma=0$ corresponds to the dissipation-less regime, while in kicked systems $\gamma=0$ stands for the maximum 
damping.

\section{Associating a classical chaotic attractor to each quantum chaotic one}
\label{sec3}

\subsection{DMKRM}

The main tools used for our investigation are the Liouville and Husimi \cite{Husimi} limiting distributions of our systems. 
For the DMKRM we have evolved $10^4$ random initial conditions in the $p \in [-\pi;\pi]$ band of the 
cylindrical phase space in the classical case (unless otherwise mentioned). For the quantum version we use the Husimi distribution of the 
evolved initial density matrix corresponding to these classical initial conditions. 
A very simple way to see the chaoticity or simplicity of these sets is by means of 
the participation ratio $\eta=(\sum_iP(p_i)^2)^{-1}/N$. This measure has its origin as a good indicator 
of the fraction of basis elements that effectively expand the quantum state. We have extended it to the classical 
case. The corresponding $\eta$ is calculated by taking a discretized $p$ distribution after $5000$ time steps, which we 
have verified is enough to reach a reasonable convergence. The number of bins is given by the Hilbert space dimension 
used in the quantum calculations, in our case is $N=3^6$. It is clear that a finer coarse-graining would slightly
change the classical $\eta$ distributions but this will not affect their main properties. This is because 
the distance among points of the simple limit cycles is almost always greater than the chosen bin size. 
The quantum equilibrium distribution is obtained in a few periods, and we have taken $50$.

\begin{figure}[htp]
\includegraphics[width=0.47\textwidth]{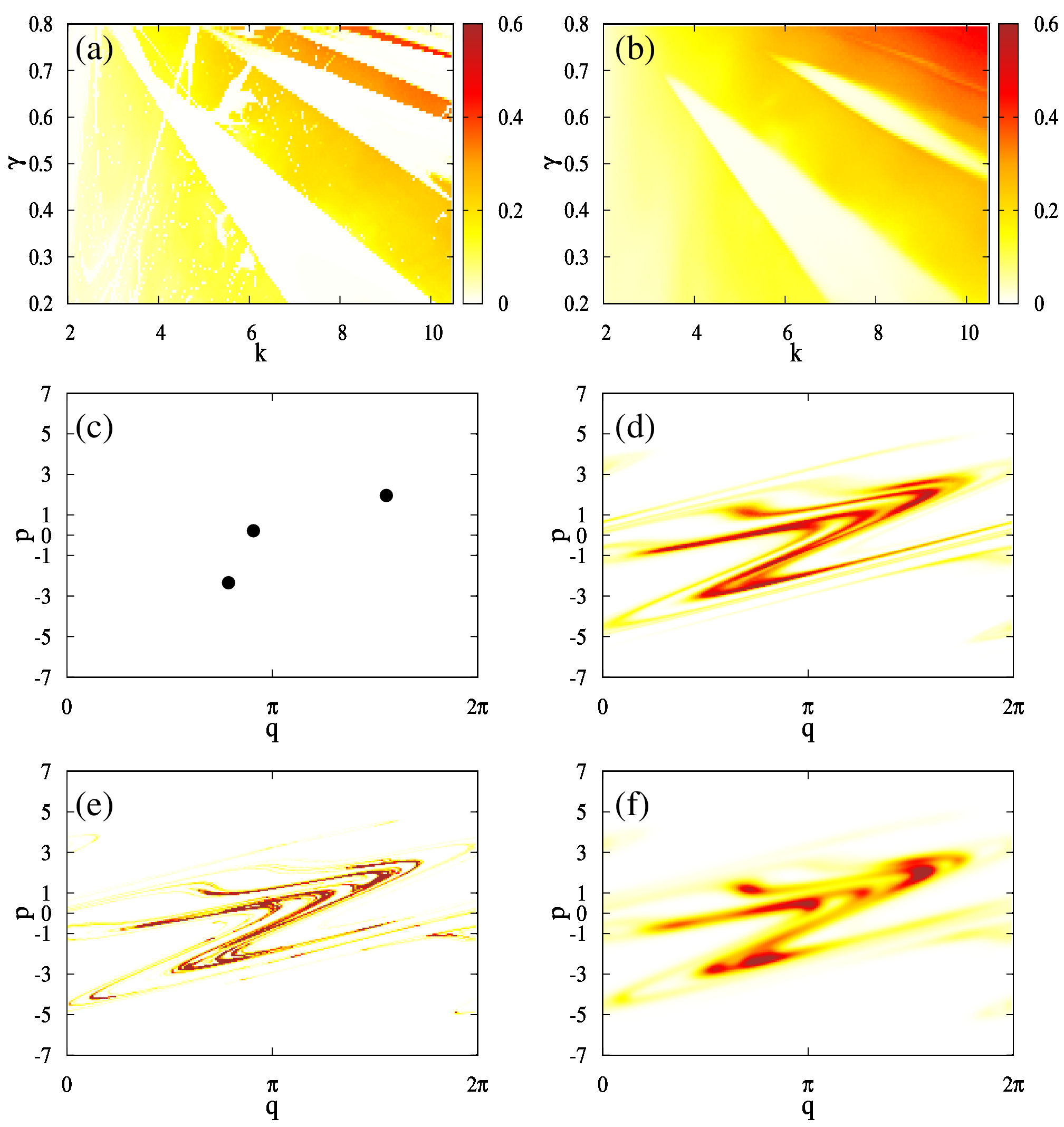} 
 \caption{(color online) In panel (a) we display the participation ratio $\eta$ for the noiseless classical DMKRM. In panel (b) we show 
 the same quantity when we add Gaussian noise of the size of $\hbar_{\rm eff}=0.019$. In (c) we show the limit cycle of period three that 
 dominates the phase space at $k=2.6$ and $\gamma=0.7$, while in (d) the classical distribution obtained by adding noise 
 corresponding to $\hbar_{\rm eff}=0.019$. In (e) we show a classical chaotic attractor (noiseless system) that is found by moving 
 in the $k$ direction ($\gamma$ fixed) for $k=2.49$ and $\gamma=0.7$. In (f) we display the quantum chaotic attractor found 
 at $k=2.6$ and $\gamma=0.7$, for $\hbar_{\rm eff}=0.019$.
 }
 \label{fig1}
\end{figure}

We have explored the parameter space of the DMKRM in a relevant region where many regular isoperiodic stable structures 
(ISSs \cite{Celestino, Carlo, Ermann, Carlo2, Beims}, originally termed as periodicity hubs \cite{Gallas}) appear. 
They are characterized by low values of $\eta$ and 
can be noticed as the clear areas with sharp borders in Fig. \ref{fig1}(a). When adding a uniform Gaussian noise of 
size $\hbar_{\rm eff}=0.019$, the resulting 
$\eta$ can be observed in Fig. \ref{fig1}(b). If we accept that this is a good measure of the quantum behaviour, 
it is clear that chaoticity is the rule while very few parameter sets correspond to point like structures. 
When noise of the size of $\hbar_{\rm eff}$ is added 
the parameter space suffers a deep transition and just a couple of regions associated to the largest of these 
ISSs keep their simplicity. 
This behaviour could seem paradoxical, and (for low dissipation values) apparently be of strictly 
quantum nature, implying that {\em purely} quantum chaotic attractors could exist without classical counterparts. 
But, can the quantum behaviour be completely disconnected from the classical behaviour 
of surrounding structures in parameter space in some cases? The explorations 
are usually done with the aid of a typical bifurcation diagram but if we see the whole parameter space things become 
clearer. 
For a representative example of an apparently purely 
quantum chaotic attractor we take $k=2.6$ and $\gamma=0.7$.
In this case the phase space is dominated by a period three limit cycle which is shown in Fig. \ref{fig1}(c) 
(we take $q=\mod{x,2\pi}$). If 
we add noise of size $\hbar_{\rm eff}=0.019$ a classical chaotic attractor develops, 
induced precisely by it and displayed in Fig. \ref{fig1}(d) (in this case we have considered $10^6$ random initial 
conditions with $p \in [-\pi;\pi]$, and accumulated their evolution over the last $100$ periods, from a total of $5000$). 
The similarity with the quantum distribution shown in Fig. \ref{fig1}(f) is remarkable. But this is not all, 
if we explore the parameter space in the orthogonal direction to $\gamma$ (which is the one typically explored 
in bifurcation diagrams), we notice that classical chaotic regions are very near, as it is the case for the 
overwhelming majority of regular regions that are embedded in them. In this case, for $k=2.49$ and $\gamma=0.7$ 
we have found a classical chaotic attractor that closely resembles the previous chaotic distributions (both 
classical with noise and quantum) (see Fig. \ref{fig1}(e)). We define the overlap 
$O=\iint {\cal D}_{\rm 1}(x,p) {\cal D}_{\rm 2}(x,p) dx dp$, where ${\cal D}_{\rm 1}(x,p)$ and ${\cal D}_{\rm 2}(x,p)$ 
are normalized phase space distributions with the same discretization. We have calculated $O$ between the classical 
distribution with noise of Fig. \ref{fig1}(d) and both, the quantum distribution 
of Fig. \ref{fig1}(f) obtaining $O=0.927$, and the quantum chaotic attractor corresponding to the classical 
one of Fig. \ref{fig1}(e) obtaining $O=0.905$.

\begin{figure}[htp]
\includegraphics[width=0.4\textwidth]{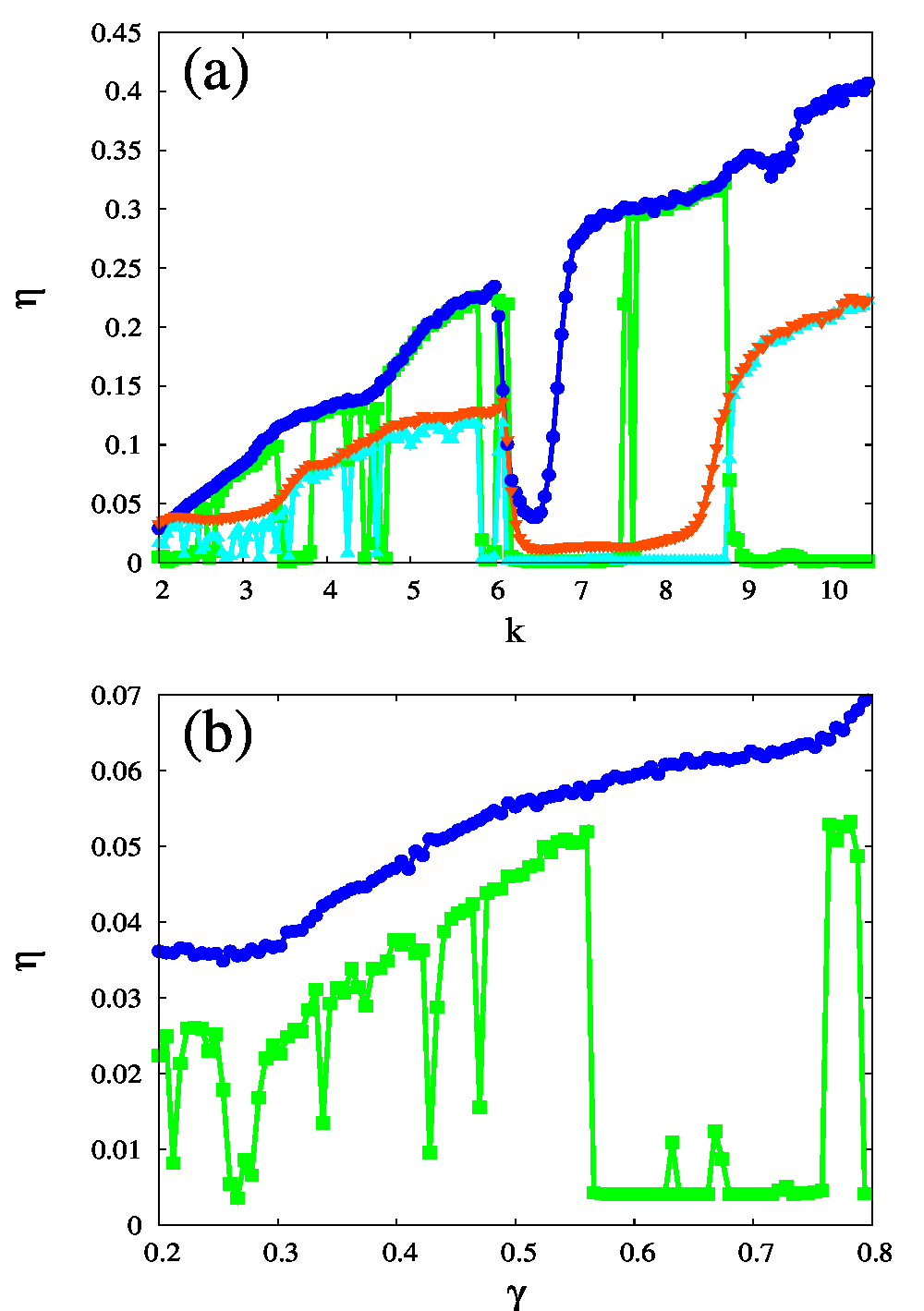} 
 \caption{(color online) Panel (a) shows $\eta$ for the DMKRM as a function of $k$. (Green) lines with squares and (blue) lines with circles 
 correspond to $\gamma=0.7$. (Cyan) lines with up triangles and (red) lines with 
 down triangles correspond to $\gamma=0.3$. Panel (b) shows $\eta$ as a function of $\gamma$. 
 (Green) lines with squares and (blue) lines with circles correspond to $k=2.6$. Each pair of lines correspond to 
 the DMKRM without and with Gaussian noise ($\hbar_{\rm eff}=0.019$), respectively.}
 \label{fig2}
\end{figure}

In Fig. \ref{fig2} we show the details of $\eta$ for different lines along $k$ and $\gamma$, in order to give 
a more precise explanation of the previous argument. If we move from $k=2.6$, while keeping $\gamma=0.7$ we 
can see that the chaoticity of distributions grows. In particular, it is the kind of 
dynamics that dominates for greater $k$ up to $k \simeq 3.5$, but also for some lower $k$ values (see Fig. \ref{fig2}(a)). 
Moreover, if we add Gaussian noise of size $\hbar_{\rm eff}=0.019$ 
the intermittency gets washed out as can be noticed by the monotonicity of the (blue) line with circles, with the 
only exception of the largest regular regions. 
For comparison purposes we also show the $\gamma=0.3$ case where again, the exceptionally 
large regular region fails to be completely smoothed out by the Gaussian fluctuations as can be seen from the deep 
fall in the (red) line with down triangles. We now come to Fig. \ref{fig2}(b) where we explore the parameter space in 
the direction of $\gamma$. If we fix $k=2.6$ and go from $\gamma=0.7$ to lower values, which means increasing the 
coupling with the environment or equivalently increasing dissipation, we have a long way to go before arriving to 
a chaotic region. This is shown by the (green) line with squares. By adding noise we recover 
the quantum behaviour without the need of changing any of the two parameters, as marked by the (blue) line with circles, which 
follows the largest $\eta$ values associated to the chaotic background. 

\subsection{DPDDS}

We extend the previous results to a generic dissipative system. For that purpose we study the classical and 
quantum limiting distributions in a stroboscopic surface of section taken at integer multiples of one period 
of the forcing. For the DPDDS we have evolved $100$ random initial conditions in the $p \in [-\pi;\pi]$ band 
(we take $p=m \dot{x}$) of the 
cylindrical phase space up to $1500$ periods, in the classical case. The Liouville distributions were obtained by 
accumulating the points of the last $50$ periods (unless otherwise noted). This assured a reasonable convergence for this systems that 
is more numerically demanding to solve than the DMKRM. Again, for the quantum version we use the Husimi distribution of the 
evolved initial density matrix corresponding to these classical initial conditions. In this case, the equilibrium distribution 
is obtained within a few periods (we have taken $50$).

\begin{figure}[htp]
\includegraphics[width=0.47\textwidth]{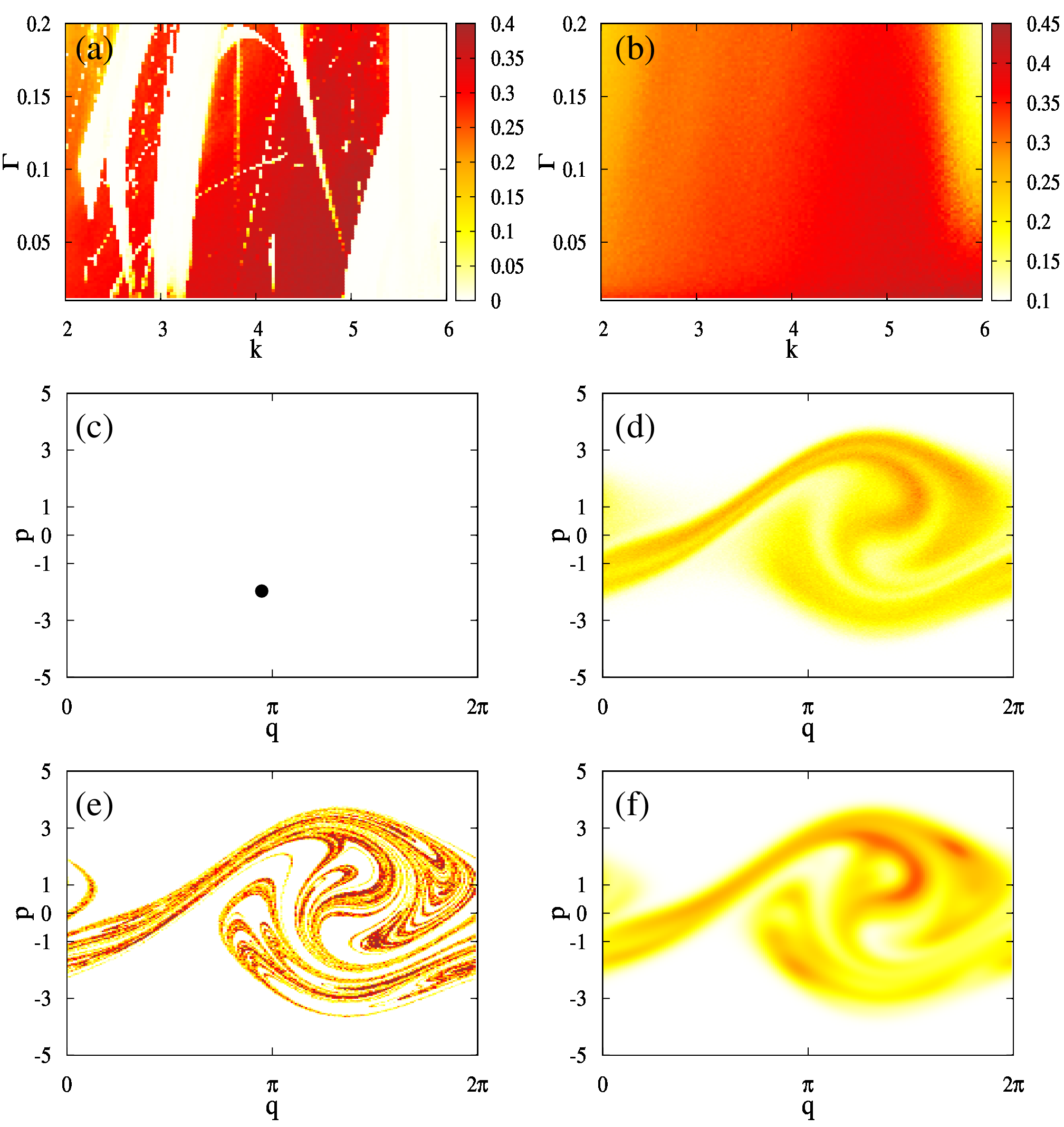} 
 \caption{(color online) In panel (a) we display the participation ratio $\eta$ for the noiseless classical DPDDS. In panel (b) we show 
 the same quantity when we add Gaussian noise of the size of $\hbar_{\rm eff}=0.041$. In (c) we show the limit cycle of period one that 
 dominates the phase space at $k=2.6$ and $\Gamma=0.06$, while in (d) the classical distribution obtained by adding Gaussian noise 
 corresponding to $\hbar_{\rm eff}=0.041$. In (e) we show a classical chaotic attractor (noiseless system) that is found by moving 
 in the $k$ direction ($\Gamma$ fixed) for $k=2.75$ and $\Gamma=0.06$. In (f) we display the quantum chaotic attractor found 
 at $k=2.6$ and $\Gamma=0.06$, for $\hbar_{\rm eff}=0.041$.
 }
 \label{fig3}
\end{figure}

In Fig. \ref{fig3}(a), it can already be seen that the morphology of the parameter space is quite similar to the one of the DMKRM. 
In fact, we can identify ISSs all over it, with their typical {\em antenna} like features which give them the {\em shrimps} familiar 
name. We can also find a larger regular structure to the right side that is the only partially surviving one 
when adding noise of the size of $\hbar_{\rm eff}=0.041$ (see Fig. \ref{fig3}(b)). In Fig. \ref{fig3}(c) we 
show a period one limit cycle that could be a representative case for a possible purely quantum chaotic attractor 
located at the low dissipation region, i.e. for $k=2.6$ and $\Gamma=0.06$ (remember that in the DPDDS the dissipation 
parameter reaches the conservative limit when $\Gamma \to 0$). If we look at Fig. \ref{fig3}(d) showing the classical 
distribution obtained by adding noise (we have used $10^4$ random initial conditions with $p \in [-\pi;\pi]$ and have 
accumulated their evolution over the last $500$ periods), it becomes clear again that this one is very much alike to that 
found for the corresponding quantum case which is shown in Fig. \ref{fig3}(f). Moreover, we are able 
to find a very similarly looking classical chaotic attractor at $k=2.75$ and $\Gamma=0.06$, displayed in Fig. \ref{fig3}(e). 
The overlap between the classical distribution with noise of Fig. \ref{fig3}(d) and the quantum distribution 
of Fig. \ref{fig3}(f) is $O=0.983$, and with the quantum chaotic attractor corresponding to the classical 
one of Fig. \ref{fig3}(e) is $O=0.976$.

\begin{figure}[htp]
\includegraphics[width=0.4\textwidth]{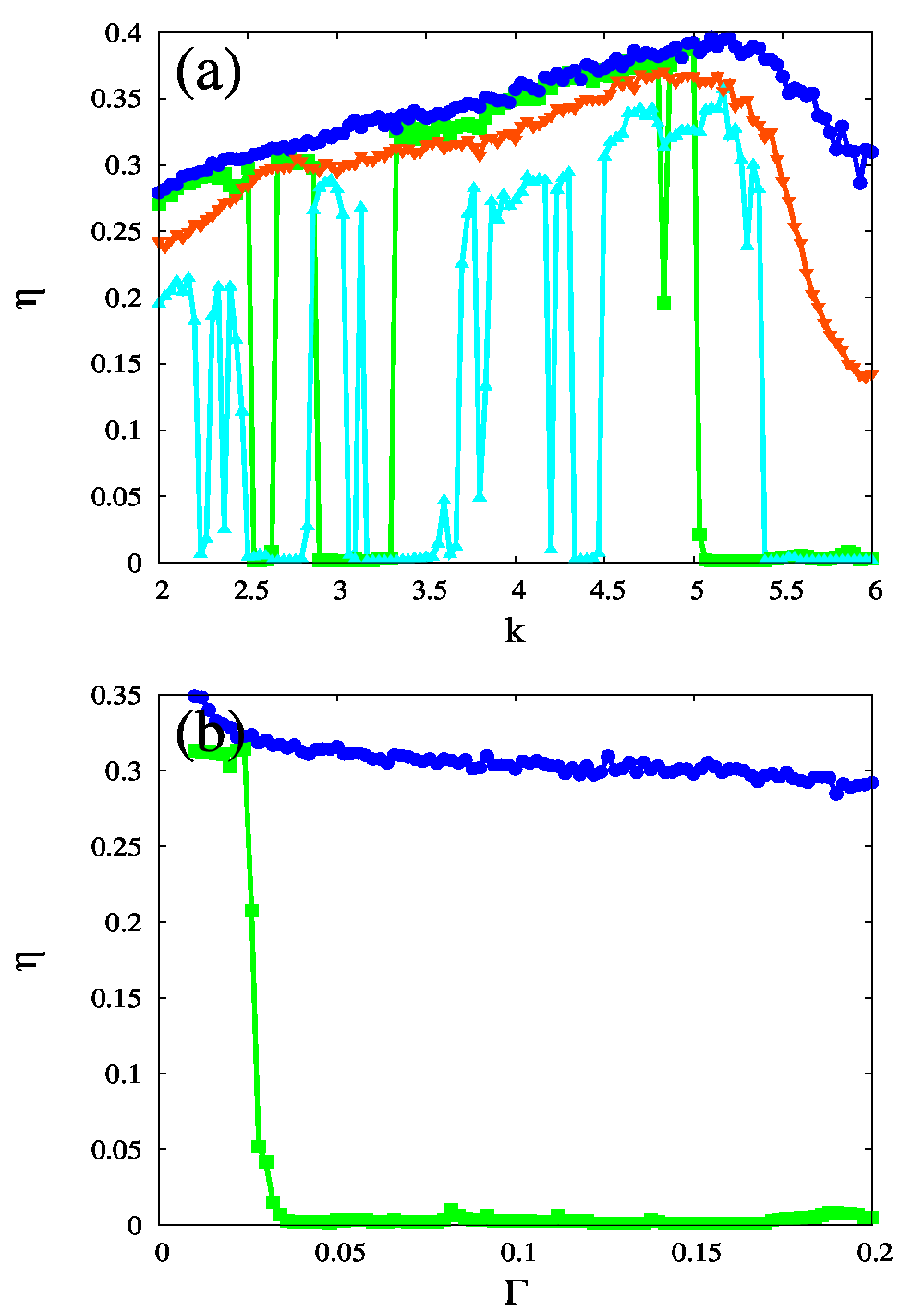} 
 \caption{(color online) Panel (a) shows $\eta$ for the DPDDS as a function of $k$. (Green) lines with squares and (blue) lines with circles 
 correspond to $\Gamma=0.06$. (Cyan) lines with up triangles and (red) lines with down triangles correspond to 
 $\Gamma=0.18$. Panel (b) shows $\eta$ as a function of $\Gamma$. (Green) lines with squares and (blue) lines with circles 
 correspond to $k=2.6$. Each pair of lines correspond to 
 the DPDDS without and with Gaussian noise ($\hbar_{\rm eff}=0.041$), respectively.}
 \label{fig4}
\end{figure}

The same detailed study of what happens when we explore the parameter space in its two main directions is applicable 
to this generic system. For example, if we look at Fig. \ref{fig4}(a) we can realize that by going up in the forcing 
strength $k$ (keeping $\Gamma=0.06$) will suffice to find a chaotic region beginning approximately at $k=2.7$. This can be seen from the sharp 
rise in the (green) line with squares. If we add noise the curve closely follows the largest values corresponding 
to the chaotic regions, so again we do not need to change any parameter to find the corresponding classical analogue of 
the quantum chaotic attractors. For comparison we show the $\Gamma=0.18$ case, where the noise only fails to rise the curve 
at the higher values of $k$ where the largest regular region lies. If we explore in the $\Gamma$ direction, departing from 
$\Gamma=0.06$ (and fixing $k=2.6$) it would take a considerable variation to reach the nearest chaotic region. This can be seen with the aid 
of the (green) line with squares in Fig. \ref{fig4}(b), which in this case happens at approximately $\Gamma=0.03$. By adding 
noise we see again how the whole curve rises (see the (blue) line with circles in Fig. \ref{fig4}(b)).

In these two subsections we have explored the main directions, the ones corresponding to the system parameters 
$k$ and $\gamma$. 
But the general way to find a classical corresponding chaotic attractor for a given quantum one would be 
to follow the shortest overall variation of both parameters in order to reach the chaotic background. 
Finally, we have verified this same behaviour for several points in parameter space, the ones shown 
are just representative cases.

\subsection{Non chaotic point like structures}

\begin{figure}[htp]
\includegraphics[width=0.47\textwidth]{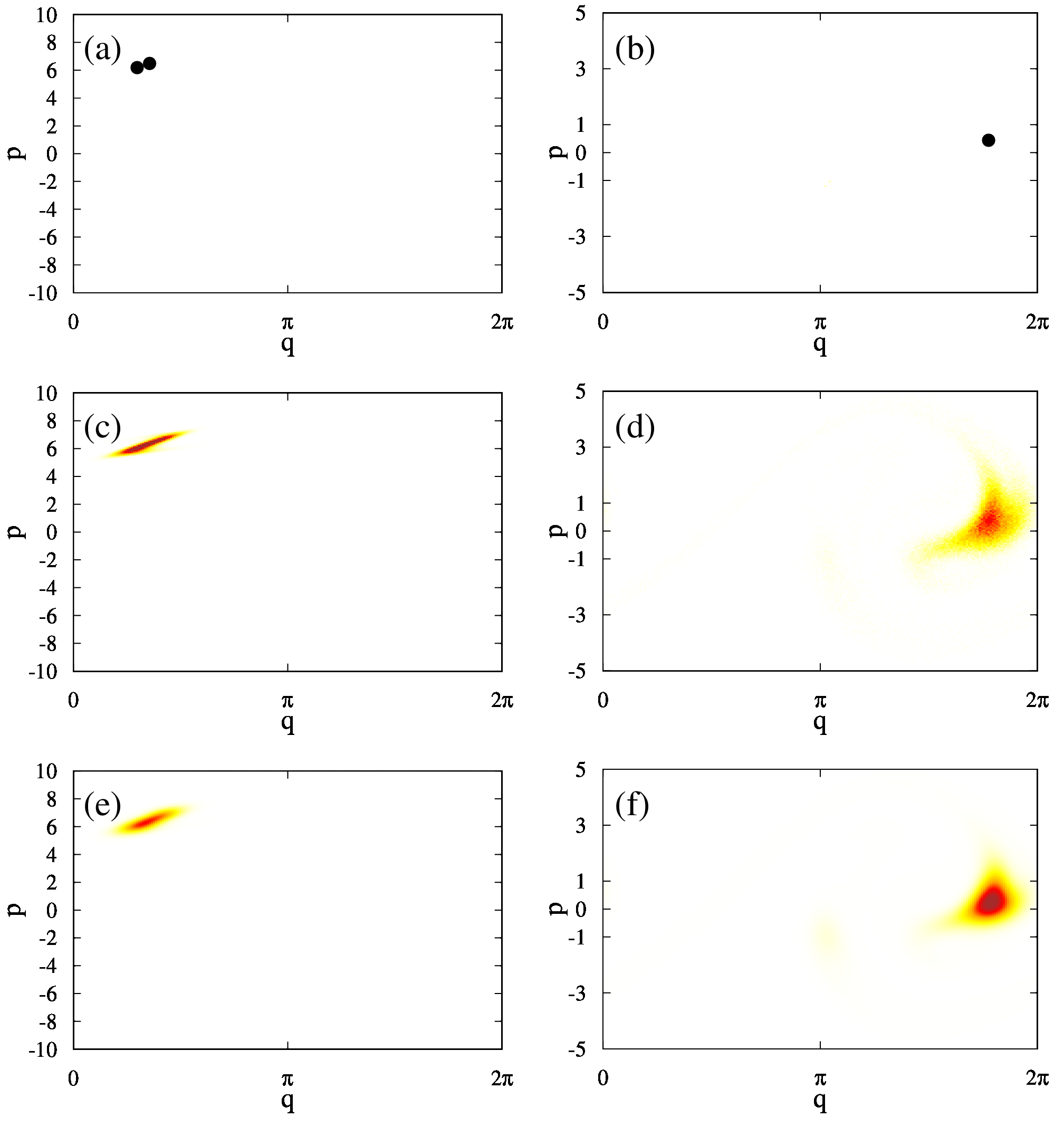} 
 \caption{(color online) In the left column results correspond to the DMKRM and in the right one to the DPDDS. 
 Panel (a) shows the period two limit cycle that dominates the phase space for $k=7.2$ and $\gamma=0.3$, 
 panel (c) the classical limiting distribution obtained with a Gaussian noise of size $\hbar_{\rm eff}=0.027$, 
 and panel (e) the quantum corresponding one. 
 Panel (b) shows the period one limit cycle that dominates the phase space for $k=6.0$ and $\Gamma=0.18$, 
 panel (d) the classical limiting distribution obtained with a noise of size $\hbar_{\rm eff}=0.041$, 
 and panel (f) the quantum corresponding one.
}
 \label{fig5}
\end{figure}

The other possibility to find a purely quantum strange attractor would be to look for the biggest ISSs which 
can have domains quite far from the chaotic background. But when the chaos is far away, what do the quantum 
distributions look like? In the left column of Fig. \ref{fig5} we analyze one representative example for 
the DMKRM and in the right one we do the same for the DPDDS (we have verified that the large regular region 
to which this case belongs extends beyond $k=6$). 

The period two limit cycle of Fig. \ref{fig5}(a) transforms into an approximately squeezed Gaussian state shown 
in Fig. \ref{fig5}(c) for the classical DMKRM with Gaussian noise, and in Fig. \ref{fig5}(e) for the quantum DMKRM. 
It is important to notice that though the quantum distribution is not a point (or two), 
this behaviour is different from the one shown in the previous two subsections. These are the simplest, point like 
quantum structures that can be found in this system and do not qualify as purely quantum attractors; 
they should be associated to simple limit cycles instead (with quantum uncertainty, of course). Interestingly, 
the same happens for the DPDDS, where the period one limit cycle shown in Fig. \ref{fig5}(b) undergoes the same 
transition to an approximately squeezed Gaussian state both for the classical with Gaussian noise and quantum DPDDS 
(see Figs. \ref{fig5}(d) and (f), respectively).

By looking at Figs. \ref{fig2}(a) and \ref{fig4}(a) we can see that these points of the parameter space 
correspond to the lowest $\eta$ values for the classical systems with noise.

\section{Conclusions}
\label{sec4}

Recently, there has been much attention directed towards the properties of the quantum to classical transition 
in dissipative systems. In particular, the study of the effects of the coupling details on the chaotic behaviour 
\cite{Eastman} and of puzzling results in optomechanics \cite{Grebogi} have provided with very interesting advances. 
In this context, quantum chaotic attractors with apparently no classical counterpart have been found in the open 
dissipative quantum Duffing system \cite{Pokharel}. On the other hand, despite known discrepancies \cite{Carlo,Carlo2} 
for some limited cases and surviving quantum effects, effective classical maps with Gaussian noise have been proposed 
as a direct replacement to obtain the main features of quantum dissipative systems. 
Important consequences have been derived from this identification \cite{Beims,Carlo3}.

We have studied two paradigmatic systems, namely the DMKRM \cite{qdisratchets}, and a generic, continuously driven DPDDS \cite{isomerization} 
We have found that we can always identify a classical chaotic attractor which corresponds to the quantum one. In 
general, there are no paradoxes in the quantum to classical correspondence of dissipative systems when we add Gaussian 
fluctuations to the classical counterparts. 
When chaotic regions are sufficiently far away from a given regular one the quantum mechanical attractor is also 
regular and the distributions become point like (with quantum uncertainty). Any quantum attractor can be explained with the help of 
these two mechanisms. This includes cases where there are coexisting attractors. Given a quantum chaotic attractor as 
the quantum version of an ISS, the general way to find the corresponding classical chaotic one consists of varying 
the parameters along the shortest way in order to reach the nearest chaotic region.

Finally, by analyzing the DPDDS we have been able to extend the validity of our method to find corresponding chaotic attractors 
and the general correspondence via Gaussian fluctuations to systems where 
even a semiclassical approximation is hard to obtain. We think that these are generic 
correspondence properties of dissipative systems \cite{Carlo3} that could have many different applications such as in many body, 
optomechanical, and reservoir engineering studies, for example. 
The influence of coherence and of different noise distributions on the quantum to classical 
correspondence will be the focus of future studies.

\section*{Acknowledgments}

Support from CONICET under project PIP 112 201101 00703 is gratefully acknowledged. D.P. acknowledges support from 
Singapore Ministry of Education, Singapore Academic Research Fund Tier-I (project SUTDT12015005).

\vspace{3pc}


\end{document}